\documentclass[epj,nopacs]{svjour}
\usepackage{graphicx}
\usepackage{amssymb}

\renewcommand{\bar}[1]{\overline{#1}}

\providecommand{\Journal}[4] {#1{\bf #2}, #3 (#4)}
\providecommand{\EPJC}{Eur. Phys. J. C }%
\providecommand{\PLB}{Phys. Lett. B } %
\providecommand{\PRL}{Phys. Rev. Lett. } %
\providecommand{\PRD}{Phys. Rev. D } %
\providecommand{\ZPC}{Z. Phys. C } %
\providecommand{\JHEP}{J. High Energy Phys. } %
\providecommand{\JPG}{J. Phys. G }

\newlength{\colw}
\begin{document}
\bibliographystyle{h-physrev4}
\title{The influence of direct $D$-meson production to the determination on the nucleon strangeness asymmetry
via dimuon events in neutrino experiments}

\author{Puze Gao\inst{1,2} \and Bo-Qiang Ma\inst{3}\thanks{Email address: mabq@phy.pku.edu.cn}}

%
%
\institute{Kavli Institute for Theoretical Physics China and
Institute of Theoretical Physics, CAS, Beijing 100190, China \and
Institute of High Energy Physics and Theoretical Physics Center for
Science Facilities, CAS, Beijing 100049, China \and School of
Physics and State Key Laboratory of Nuclear Physics and Technology,
Peking University, Beijing 100871, China }

\date{Received: date / Revised version: date}
%
\abstract{Experimentally, the production of oppositely charged
dimuon events by neutrino and anti-neutrino deep inelastic
scattering (DIS) is used to determine the strangeness asymmetry
inside a nucleon. Here we point out that the direct production of
$D$-meson in DIS may make substantial influence to the measurement
of nucleon strange distributions. The direct $D$-meson production is
via the heavy quark recombination (HQR) and via the light quark
fragmentation from perturbative QCD (LQF-P). To see the influence
precisely, we compute the direct $D$-meson productions via HQR and
LQF-P quantitatively and estimate their corrections to the analysis
of the strangeness asymmetry. The results show that HQR has stronger
effect than LQF-P does, and the former may influence the
experimental determination of the nucleon strangeness asymmetry.
\PACS{ 
     } 
} 
%


\titlerunning{The leading particle effect from light quark fragmentation
in charm hadroproduction} \maketitle


\section{Introduction}
\label{intro}

Studying strange and anti-strange quark distributions of a nucleon
is an important part in the study of the nucleon structure. An
asymmetric strange distribution, i.e., the parton distribution
function (PDF) of strange quark being not equal to that of
anti-strange quark inside a nucleon, is naturally predicted by some
non-perturbative models~\cite{bm96,sig87,bur92}. Further clear check
of the strangeness asymmetry is not only important for the study of
the nucleon structure itself, but also for understanding relevant
phenomenon in some experiments. For example, the so-called NuTeV
anomaly phenomenon~\cite{NT1,NT2} can be explained by nonzero
strangeness asymmetry~\cite{oln03,kre04,dm04,alw04,dxm04,wak04}.

Because of the smallness of the strange and anti-strange components
in a nucleon, to measure the strangeness asymmetry is a challenging
job indeed. The most sensitive reaction to measure the strange and
anti-strange distributions is the production of dimuon events in
neutrino and anti-neutrino nucleon deep inelastic scattering (DIS).
To leading order (LO) of the dimuon production, the events are
caused by the charged-current (CC) charm production subprocesses
$\nu_\mu +s\;(d) \rightarrow \mu^- +c$ or $\bar{\nu}_\mu
+\bar{s}\;(\bar{d}) \rightarrow \mu^+ +\bar{c}$ and a cascade decay
$c\to \mu^+ +\cdots$ or $\bar{c}\to \mu^- +\cdots$. The relevant
transition $\nu_\mu(\bar{\nu}_\mu) + d(\bar{d}) \rightarrow
\mu^-(\mu^+) +c(\bar{c})$ is Cabibbo suppressed whereas
$\nu_\mu({\bar\nu}_\mu) +s(\bar{s}) \rightarrow \mu^-(\mu^+)
+c(\bar{c})$ is Cabibbo favored. Thus the dimuon events are
sensitive to the strange and anti-strange distributions of the
target nucleon. In the literature, measurements of the strangeness
asymmetry via oppositely signed dimuon are carried out by CCFR and
NuTeV~\cite{CCFR93,CCFR95,NTdimu01,mason04,mason06,mason07}
experiments.

The results of global analysis~\cite{bar00,oln03,lai07} indicate the
strangeness asymmetry $S^-\equiv\int \xi[s(\xi)-\bar s(\xi)]d\xi$
likely to be positive, e.g., in Ref.\cite{lai07} $-0.001<S^-<0.005$
is obtained. The recent NuTeV reanalysis up-to next-to-leading order
(NLO) of perturbative QCD (pQCD) with improved method supports the
positive strangeness asymmetry $S^-=0.00196\pm
0.00046({\mathrm{stat}}) \pm
0.00045({\mathrm{syst}})^{+0.00148}_{-0.00107}\\({\mathrm{external}})$~\cite{mason07},
which is consistent with the global analysis, although at early
stage the analysis of CCFR and NuTeV dimuon events at LO and even
NLO do not support the strangeness asymmetry
\cite{CCFR95,NTdimu01,mason04}.

In this work, we take a systematic study on the influence of direct
$D$-meson production at order $\alpha_s^2$ to the determination of
the nucleon strangeness asymmetry via dimuon events in neutrino
experiments. With consideration of the experimental kinematic cuts
in CCFR and NuTeV, we point out that there are two kinds of direct
$D$-meson production: heavy quark recombination process (HQR) and
light quark fragmentation in the pQCD picture (LQF-P), which can
contribute to the cross section difference between neutrino and
anti-neutrino induced CC DIS.  The direct $D$-meson production may
influence the strangeness measurement (determination) that depends
on their magnitude, so we further calculate the production
quantitatively and investigate their influence to the strangeness
determination. Although our preliminary result on HQR process has
been briefly reported in Ref.\cite{gm07hqr}, in this paper, we would
present it in more detail and with some improvements. From final
results we find that the influence of the direct $D$-meson
production to the measurement of the strangeness asymmetry could not
be negligible.

The rest of the paper is organized as follows. In section II, we
discuss CC charm production to the dimuon cross sections at LO and
NLO. And then we present the two kinds of direct $D$-meson
production: HQR and LQF-P, which in fact are of high order ones, and
show how the two processes can affect the extraction of the nucleon
strangeness asymmetry. In section III, we show the numerical
calculation about HQR direct production of $D$-meson and estimate
the influence due to the direct production. In section IV, we show
numerical calculation and estimate as in previous section about
those for LQF-P production. In conclusion section (section V), we
summarize and discuss the obtained results.

\section{Dimuon events and direct $D$-meson production in neutrino DIS}

Experimentally with the CCFR and NuTeV detectors, the oppositely
charged dimuon signal induced by CC charm production in $\nu_\mu$
($\bar\nu_\mu$) DIS has a distinct feature and is not very difficult
to be detected. The first muon of the dimuon is from the
$\nu_\mu(\bar\nu_\mu)$ vertex, and the second muon is from a little
delayed muonic decay of the produced charm. The life time of $\pi$
and $K$ is much longer than charmed, so those muons from $\pi$ or
$K$ meson decay can be largely eliminated, i.e., they will not
contribute to the dimuon events concerned here.

According to pQCD factorization theorem, for $\nu_\mu$-proton DIS,
to LO the differential cross section for dimuon production induced
by CC production of charmed hadron $H$ can be expressed
as~\cite{gm05}:
\begin{eqnarray}%
{d^3\sigma_{\nu_\mu P\rightarrow\mu^-\mu^+X}\over d\xi dQ^2dz}
=&&{G_F^2\over \pi r_w^2}f_c \left[d(\xi,Q^2)|V_{cd}|^2 +
s(\xi,Q^2)|V_{cs}|^2\right]\nonumber\\
&&\times\sum_H D_c^H(z)Br_H \,.\label{nu-P-dimu}
\end{eqnarray}
where $d(\xi,Q^2)$ and $s(\xi,Q^2)$ are the parton distribution
functions (PDFs) of $d$ and $s$ quarks in the proton, and $\xi$,
relating to the Bjorken scaling variable $x$ through $\xi=
x(1+m_c^2/Q^2)$, is the light-cone momentum fraction of the struck
quark; $Q^2=-(p-k)^2$ is the minus squared invariant momentum
transfer with $p$ and $k$ being the momentum of the incident
$\nu_\mu$ and the scattered $\mu^-$ respectively. $r_w\equiv
1+Q^2/M_W^2$ and $f_c\equiv 1-{m_c^2/S\xi}$ with $M_W$ being the
W-boson mass and $S$ being the squared C.M. energry of the
neutrio-proton system; $D_c^H(z)$ is the fragmentation function for
a charm quark to the charmed hadron $H$, and $Br_H$ is the branching
ratio of muonic decay for $H$. Carrying out the integration over $z$
and the summation over $H$, with the definition $B_{ef}\equiv \int
dz\sum_H D_c^H(z)Br_H$, for target nucleus with proton number $P$
and neutron number $N$, the differential cross section can be
expressed as
\begin{eqnarray}%
{d^2\sigma_{\nu_\mu A\rightarrow\mu^-\mu^+X}\over d\xi dQ^2}
=&&{G_F^2\over \pi r_w^2}f_c \Big[{P\,d(\xi,Q^2)+N\,u(\xi,Q^2)\over
P+N}|V_{cd}|^2 \nonumber\\
&&+ s(\xi,Q^2)|V_{cs}|^2\Big]B_{ef} \,.\label{nu-dimu}
\end{eqnarray}
where the PDFs in the
neutron is related to PDFs of proton by $d_n(\xi)=u(\xi)$, 
$s_n(\xi)=s(\xi)$ etc.

Since $|V_{cd}|^2\sim 0.05$ and $|V_{cs}|^2\sim 0.9$~\cite{PDG06},
the $\nu_\mu$ induced dimuon cross section is sensitive to the
strange distribution $s(\xi,Q^2)$ in the target nucleus. Similarly,
$\bar\nu_\mu$ induced dimuon cross section is sensitive to
anti-strange distribution $\bar s(\xi,Q^2)$. The difference between
dimuon cross sections induced by $\nu_\mu$ and $\bar\nu_\mu$ is
directly related to the strange distribution asymmetry. The
difference to LO can be expressed as:
\begin{eqnarray}\label{eq:SALO}
&&{d^2\sigma_{\nu_{\mu}N\rightarrow \mu^- \mu^+ X}\over d\xi dQ^2 }-
{d^2\sigma_{\bar\nu_{\mu}N\rightarrow \mu^+ \mu^- X}\over d\xi dQ^2}
\nonumber\\
=&&{G_F^2 \over \pi r_w^2}f_cB_{ef} \Big\{
\big[s(\xi,Q^2)-\bar s(\xi,Q^2)\big ]|V_{cs}|^2\nonumber\\
&&+{1\over P+N}\big [P\,d_v(\xi,Q^2)+N\,u_v(\xi,Q^2)\big
]|V_{cd}|^2\Big\}\,,
\end{eqnarray}
where $d_v(\xi,Q^2)=d(\xi,Q^2)-\bar d(\xi,Q^2)$ and
$u_v(\xi,Q^2)=u(\xi,Q^2)-\bar u(\xi,Q^2)$ are valence distributions
of proton.

The Feynman diagrams for CC charm production at NLO from subprocess
of $\nu_\mu N$ DIS are shown in FIG. 1~\cite{mason06}. In fact, the
last two diagrams of FIG. 1~\cite{mason06} are involved in the
leading logarithm (LL) evolution for the parton flavor-singlet
components of PDFs, and the flavor singlet components of the PDFs
contribute to the dimuon cross sections symmetrically for the
$\nu_\mu$- and $\bar\nu_\mu$-induced DIS. Thus even up-to NLO, the
difference between $\nu_\mu$- and $\bar\nu_\mu$-induced dimuon cross
section is still proportional to $[s(\xi,Q^2)-\bar
s(\xi,Q^2)]|V_{cs}|^2 +{1\over
P+N}[P\,d_v(\xi,Q^2)+N\,u_v(\xi,Q^2)]|V_{cd}|^2$, i.e., we need
consider neither the last two diagrams of FIG.1 nor the flavor
singlet components of PDFs in the dimuon cross section difference.
\begin{figure}
\includegraphics*[width=0.9\colw]{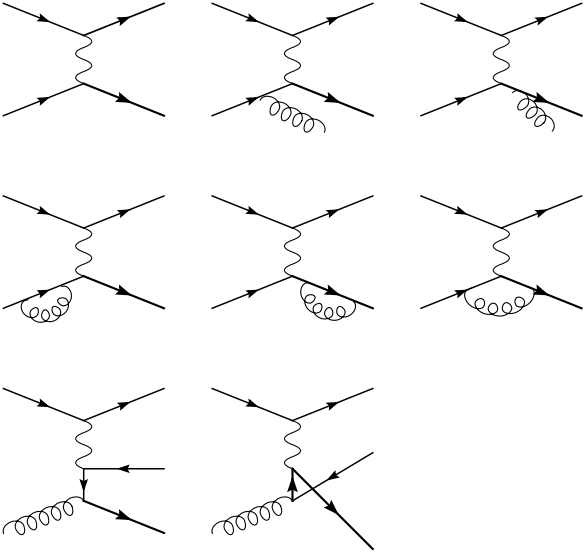}
\caption{ Feynman diagrams for calculating the charm production by
  energetic $\nu_\mu$ collision via CC weak interaction up-to NLO.
  The thick lines denote those of c or $\bar c$ quarks.} 
\end{figure}

However,
the direct $D$-meson\footnote{The $D^*$ meson, the excited
($c\bar{q}$) bound state in $^3S_1$, has very great cross section in
production to compare with the $D$-meson production, and decays into
$D$ meson via strong and/or electromagnetic interaction with almost
100\% branching ratio, thus the consequence of the $D^*$-meson
production will be as direct $D$-meson production in accounting the
muons in the dimuon events. Therefore throughout the paper
``$D$-meson production'' always mean the production of $D$ and
$D^*$.} production that convolutes to the valence components in the
nucleon, as we will discuss in the following, can raise the rate of
dimuon production after experimental kinematic cuts, therefore
determinations of strangeness asymmetry can be distorted by the
direct $D$-meson production in certain degree. 

Now let us focus the light on the contributions from the direct
$D$-meson production, although according to pQCD the lowest order
Feynman diagrams of the direct production are of the order
$\alpha_s^2$.

One of the direct $D$-meson production mechanism, the so-called
heavy quark recombination (HQR) process, at LO level is described by
the diagrams in FIG. 2. It is stimulated from the heavy quark
recombination mechanism ~\cite{br02b,br02c,br02prl}, which combines
a heavy quark and a light anti-quark of similar velocity to form a
meson. Refs.~\cite{br02c,br02prl} employ simple pQCD pictures and
explain the charm photoproduction asymmetry and the leading particle
effect~\cite{E791} successfully.
\begin{figure}
{\includegraphics[width=0.9\columnwidth]{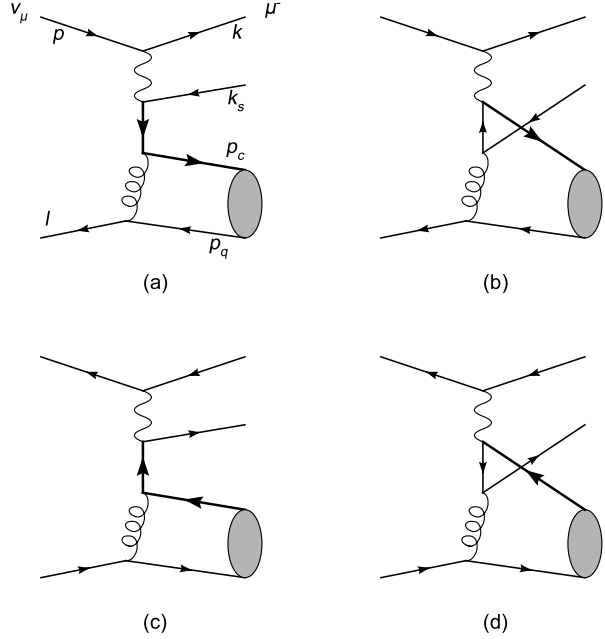}}\vspace{0.5cm}
\caption{(a) and (b) are diagrams for $(c\bar q)$ recombination into
a $D$ meson in $\nu_\mu$ induced process Eq.(\ref{eq:nuHR}); (c) and
(d) are diagrams for $(\bar c q)$ recombination into a $\bar D$
meson in $\bar\nu_\mu$ induced process Eq.(\ref{eq:nubarHR}). The
thick lines in the diagrams denote those of heavy quarks, and the
shaded blobs denote $D$ or $\bar D$ mesons.}
\end{figure}

Namely the difference between $\nu_\mu$- induced and
$\bar\nu_\mu$-induced dimuon cross sections caused by the HQR
production of direct $D$-meson can be computed:
\begin{eqnarray}\label{eq:csdHR}
&&\left [{d^2\sigma_{\nu_{\mu}N\rightarrow \mu^- \mu^+ X}\over d\xi
dQ^2 }- {d^2\sigma_{\bar\nu_{\mu}N\rightarrow \mu^+ \mu^- X}\over
d\xi dQ^2}\right]_{HR}\nonumber\\
 =&&\sum_{q,D}\int dx[\bar
q(x,\mu^2)-q(x,\mu^2)]{d^2\hat{\sigma}_{D}\over d\xi dQ^2} Br_{D},
\end{eqnarray}
where $q$ denotes a possible light quark in the target and $D$
denotes a produced $D$-meson. $d\hat\sigma_D$ denotes the
differential cross section of the subprocess
\begin{equation}\label{eq:nuHR}
\nu_\mu+\bar q\rightarrow \mu^-+\bar s\;(\bar d)+D\,,
\end{equation}
which, in terms of $CP$ transformation, is known to be equal to that
of the subprocess
\begin{equation}\label{eq:nubarHR}
\bar\nu_\mu+ q\rightarrow \mu^++ s\;(d)+\bar D\,.
\end{equation}
And $Br_D$ denotes the muonic decay rate of the $D$-meson. In fact,
from Eq.(\ref{eq:csdHR}), it is easy to realize that only the $u-d$
valance components of the PDFs contribute to the cross section
difference ($s-\bar s$ contribution comparatively negligible), thus
$q$ could be either $u$ or $d$. Such a contribution may reduce the
strange distribution asymmetry
$S^-(\xi,Q^2)\equiv\xi[s(\xi,Q^2)-\bar s(\xi,Q^2)]$ value obtained
by the analysis where the HQR process has been ignored. Thus,
roughly speaking, the true value of the strange distribution inside
a nucleon $S^-(\xi,Q^2)$ may be changed from the existent analysis
result which has ignored the HQR effects by a positive correction
factor $\delta S^-_{\mathrm{HR}}(\xi,Q^2)$
\begin{equation}\label{eq:SArl}
S^-_\mathrm{real}(\xi,Q^2)=S^-_\mathrm{analy}(\xi,Q^2)+\delta
S^-_\mathrm{HR}(\xi,Q^2).
\end{equation}

For a quantitative estimate of the HQR correction to the measured
strangeness asymmetry, we suppose that the Cabibbo suppressed
valence contribution in Eq.(\ref{eq:SALO}) has been deducted in LO
(or NLO) analysis of dimuon events though present knowledge of the
PDFs,
\begin{eqnarray}
f_{\rm{LO}}|V_{cs}|^2&&[s(\xi,Q^2)-\bar
s(\xi,Q^2)]^{\rm{LO}}_{\rm{analy}}=\Big [{d^2\sigma_{\nu_{\mu}}\over
d\xi dQ^2} -{d^2\sigma_{\bar\nu_{\mu}}\over d\xi
dQ^2}\Big]_{\rm{ex}}\nonumber\\
&&-f_{\rm{LO}}|V_{cd}|^2 {P\,d_v(\xi,Q^2)+N\,u_v(\xi,Q^2)\over P+N}.
\end{eqnarray}
Where $f_{\rm{LO}}=G_F^2f_cBr_c/\pi r_w^2$, is the coefficient in
Eq.(\ref{eq:SALO}) for LO cross section, and
${d^2\sigma_{\nu_{\mu}(\bar\nu_\mu)}\over d\xi dQ^2}$ is the
differential cross section for $\nu_\mu(\bar\nu_\mu)$-induced dimuon
production. While in fact, the measured cross section also includes
the contribution from HQR as Eq.(\ref{eq:csdHR}), which should be
deducted to obtain the real strange distribution asymmetry
\begin{eqnarray}
&&f_{\rm{LO}}|V_{cs}|^2[s(\xi,Q^2)-\bar
s(\xi,Q^2)]^{\rm{LO}}_{\rm{real}}\nonumber\\
=&&\left[{d^2\sigma_{\nu_{\mu}}\over d\xi
dQ^2}-{d^2\sigma_{\bar\nu_{\mu}}\over d\xi dQ^2}\right]_{\rm{ex}}
-\left[{d^2\sigma_{\nu_{\mu}}\over d\xi dQ^2}
-{d^2\sigma_{\bar\nu_{\mu}}\over d\xi
dQ^2}\right]_{\rm{HQR}}\nonumber\\
&&-{1\over P+N}f_{\rm{LO}}|V_{cd}|^2
{[P\,d_v(\xi,Q^2)+N\,u_v(\xi,Q^2)]}\nonumber\\
=&&f_{\rm{LO}}|V_{cs}|^2[s(\xi,Q^2)-\bar
s(\xi,Q^2)]^{\rm{LO}}_{\rm{analy}}\nonumber\\
&&-\left[{d^2\sigma_{\nu_{\mu}}\over
d\xi dQ^2}-{d^2\sigma_{\bar\nu_{\mu}}\over d\xi
dQ^2}\right]_{\rm{HQR}}
\end{eqnarray}
Thus, the real strange distribution asymmetry
$S^-(\xi,Q^2)\equiv\xi[s(\xi,Q^2)-\bar s(\xi,Q^2)]$ can be deduced.
\begin{eqnarray}
S^{-\rm{LO}}_{\rm{real}}(\xi,Q^2)&=&S^{-\rm{LO}}_{\rm{analy}}(\xi,Q^2)\nonumber\\
&&-{\xi\over f_{\rm{LO}}|V_{cs}|^2}\left
[{d^2\sigma_{\nu_{\mu}}\over d\xi dQ^2 }-
{d^2\sigma_{\bar\nu_{\mu}}\over d\xi
dQ^2}\right]_{\rm{HQR}}\,.\label{eq:Sdeduce}
\end{eqnarray}
Thus for LO analysis the HQR correction to strange distribution
asymmetry can be estimated from
Eq.(\ref{eq:csdHR},\ref{eq:SArl},\ref{eq:Sdeduce})
\begin{eqnarray}\label{eq:SAHR}
\delta S^-_\mathrm{HR}(\xi,Q^2)\simeq&&
 {\pi r_w^2\xi\over G_F^2  f_c B_{ef}|V_{cs}|^2}
\nonumber\\
\times\sum_{q,D}\int &&dx[q(x,\mu^2)-\bar
q(x,\mu^2)]{d^2\hat{\sigma}_{D}\over d\xi dQ^2} Br_{D}\,.
\end{eqnarray}
For NLO analysis, the factor $f_{\rm{LO}}$ in Eq.(\ref{eq:Sdeduce})
should be replace by a NLO coefficient, and the  HQR correction
should differ from this LO one Eq.(\ref{eq:SAHR}) by a K factor.
Numerical calculation of $\delta S^-_\mathrm{HR}(\xi,Q^2)$ will be
the content of the next section.

The other possible mechanism is the so-called light quark
fragmentation process in the order $\alpha_s^2$ of pQCD (LQF-P). Its
subprocess is described by FIG.3. From FIG.3 it is easy to realize
that LQF-P contributes not only to the inclusive oppositely charged
dimuon events but also to trimuon and the inclusive same charged
dimuon events. Here we focus only on the contribution from LQF-P to
the inclusive oppositely charged dimuon with the restriction that
the second muon comes from the produced direct $D$-meson. This
mechanism is considered as fragmentation of a light quark into a
$D$-meson in perturbative picture, which is different from the
non-perturbative picture discussed in Ref.~\cite{gm07}.
\begin{figure}
{\includegraphics[width=0.9\columnwidth]{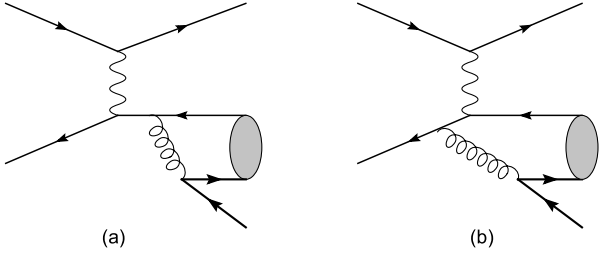}}
\caption{Diagrams for the LQF-P process in $\nu_\mu$ induced CC DIS.
The produced $D$ meson will decay partially into $\mu^+$ to form the
second muon. Thick lines denote that of heavy quarks, and shaded
blobs denote the $D$ meson produced directly.}
\end{figure}


For nucleus target with proton number $P$ and neutron number $N$,
the $\nu_\mu$-induced dimuon cross section from the LQF-P production
can be expressed as
\begin{eqnarray}\label{eq:csd-LQF-P0}
&&\left [{d^2\sigma_{\nu_{\mu}N\rightarrow \mu^- \mu^+ X}\over d\xi
dQ^2 }\right]_\mathrm{LQF-P}\nonumber\\
=&&\sum_{D}\int dx{P\,\bar u(x)+N\,\bar d(x)\over
P+N}{d\hat\sigma_{\bar
 c D}\over d\xi dQ^2}Br_D\,.
\end{eqnarray}Where $d\hat\sigma_{\bar c D}$ denotes the cross section of
subprocess $\nu_\mu+\bar u\rightarrow\mu^- + D + \bar c$ (diagrams
in FIG.3). The $\bar\nu_\mu$-induced dimuon cross section from LQF-P
production can be obtained by $CP$ transformation to that of $\nu$,
and thus the $\nu_\mu$ and $\bar\nu_\mu$-induced dimuon cross
section difference from LQF-P process can be expressed as
\begin{eqnarray}\label{eq:csd-LQF-P}
&&\left [{d^2\sigma_{\nu_{\mu}N\rightarrow \mu^- \mu^+ X}\over d\xi
dQ^2 }-{d^2\sigma_{\bar\nu_{\mu}N\rightarrow \mu^+ \mu^- X}\over
d\xi dQ^2 }\right]_\mathrm{LQF-P}\nonumber\\ =&&-\int
dx{P\,u_v(x)+N\,d_v(x)\over P+N}\sum_{D}{d\hat\sigma_{\bar c D}\over
d\xi dQ^2}Br_D.
\end{eqnarray}
Note here that to compare with the valance components, the other
components are tiny, so in Eq.(\ref{eq:csd-LQF-P}), except the
valance components, the other components are ignored.

Numerical calculation of the LQF-P process will be presented in
section IV.

\section{Calculation of the HQR process}

As pointed out in the above section, the direct $D$ meson production
from HQR process (diagrams in FIG. 2) may influence the measurement
of the nucleon strangeness asymmetry. Our calculation of the HQR
process follows the method in Refs.~\cite{br02b,br02c,br02prl}.
Namely we `factorize' the production into the two steps, one of them
(the one we start with) is the production of the relevant $^1S_0$ or
$^3S_1$ state of $(c\bar q)$, the other one (the following one) is
the combined quark pair to evolve into the relevant $D$- or
$D^*$-meson under certain possibility, that is determined by other
experiments.

When a charm $c$ quark and a $\bar q$ light anti-quark with momentum
$p_c$ and $p_q$ respectively are produced with $p_q$ to be small in
the $c$-quark rest frame, the $c$ and $\bar q $ can be constructed
into a $^1S_0$ color-singlet state $(c\bar q)$ if $q$- and $c$-quark
have the same color, then we have the following substitution in
amplitude for the combination of the partons:
\begin{equation}\label{eq:substitution}
v_j(p_q)\bar u_i(p_c)\rightarrow x_q{\delta_{ij}\over N_c}m_c
(p\!\!\!/_c-m_c)\gamma_5\,,
\end{equation}
with $p_q=x_qp_c$ and limit $x_q\rightarrow 0$ may be taken.
Accordingly, for the color-singlet $^3S_1$ state $(c\bar q)$,
$\gamma_5$-matrix in Eq.~(\ref{eq:substitution}) should be replaced
by $\epsilon\!\!\!/$, where $\epsilon^\mu(P,\lambda)$ is the
polarization vector of the $^3S_1$ $(c\bar q)$ state with total
momentum $P$ and polarization $\lambda$ of the quark pair.
$\delta_{ij}$ in Eq.(\ref{eq:substitution}) is the color factor for
color-singlet state with quark colors $i\,,j=1,2,3$. Corresponding
to the diagrams in FIG. 2(a)(b), the amplitude for the production of
the $(c\bar q)$ color-singlet $^1S_0$ state induced by $\nu_\mu$ may
be written down directly:
\begin{eqnarray}\label{eq:amplitue}
M&=&{16\pi G_F\alpha_s m_c\delta_{mn}\over 9\sqrt{2}r_w(2l\cdot
p_c)} V_{cf}\bar u(k)\gamma^\mu(1-\gamma_5)u(p)\nonumber\\
&\times&\bar u(l)\gamma^\nu(p\!\!\!/_c-m_c)\gamma_5 \Big[\gamma_\nu
{p\!\!\!/-k\!\!\!/-k\!\!\!/_s+m_c\over(p-k-k_s)^2-m_c^2}\gamma_\mu(1-\gamma_5)\nonumber\\
&&+\gamma_\mu(1-\gamma_5){l\!\!\!/-k\!\!\!/_s\over
(l-k_s)^2}\gamma_\nu \Big ]v(k_s)\,.
\end{eqnarray}
Here $p$, $l$, $k$ and $k_s$ denote the momenta of the $\nu_\mu$,
the initial incoming $\bar q$, the produced $\mu^-$ and the produced
$\bar s$ or $\bar d$ respectively. $V_{cf}$ is the CKM matrix
element with $f=s, d$. Color factor
$T_{mi}^b\delta_{ij}T^b_{jn}={4\over 3}\delta_{mn}$ has been
included in the amplitude of Eq.(\ref{eq:amplitue}). Squaring the
amplitude, averaging over the spin and color of the particles in the
initial state, and summing up the spin, color and possible flavors
($f=s,d$) of the particles in the final state, the `averaged and
summed' squared amplitude is
\begin{eqnarray}\label{eq:M21}
|\bar M|^2[(c\bar q)_{^1S_0}^{\bf 1}]&=&{4\pi^2 16^3 G_F^2\alpha_s^2
m_c^2\over 81 r_w^2
B(s-A-B)^3}\nonumber\\
&\times&\Big\{C^2[(s-A)m_c^2+(s-A-B)Q^2] \nonumber\\
&+&m_c^2C[(s-A)m_c^2-(s-A-B)(s-Q^2)] \nonumber\\
&+&m_c^2A[s(Q^2+m_c^2)-BQ^2]\Big\}.
\end{eqnarray}
Here $s\equiv(p+l)^2=Sx$ is the squared C.M. energy of the
subprocess. The variables $A$, $B$ and $C$ are defined as
$A=2(l\cdot k)$, $B=2(l\cdot k_s)$, $C=2(p\cdot k_s)$.

For the color-singlet $^3S_1$ $(c\bar q)$-state production, the
`averaged and summed' squared amplitude is
\begin{eqnarray}\label{eq:M22}
|\bar M|^2[(c\bar q)_{^3S_1}^{\bf 1}]&=&{4\pi^2 16^3 G_F^2\alpha_s^2
m_c^2\over 81 r_w^2
B(s-A-B)^3}\nonumber\\
&\times&\Big\{3C^2[(s-A)m_c^2+(s-A-B)Q^2]
\nonumber\\
&+&C\big[(3Q^2-5s+2A)(s-A-B)m_c^2\nonumber\\
&+&3(s-A)m_c^4+4(B-S)(s-A-B)Q^2\big]
\nonumber\\
&+&[s(Q^2+m_c^2)-BQ^2]\nonumber\\
&\times&[2(s-A-B)(s-B-m_c^2)-Am_c^2]\Big\}.
\end{eqnarray}

For the color-octet $(c\bar q)$ either in $^1S_0$ or $^3S_1$ state,
the amplitude differs from the `color-singlet' ones only in the
color factor, e.g., the $\delta_{mn}$ in Eq.(\ref{eq:amplitue})
should be replaced by $\frac{\sqrt N_c}{4\sqrt 2}T_{mn}^a$ with
$a=1,2,\cdots,8$.

Then the produced $(c\bar q)$-states will evolve into either $D$ or
$D^*$ meson with certain probabilities. The cross section for the
subprocess of a $D$-meson production $\nu_\mu+\bar
q\rightarrow\mu^-+D+\bar s\,(\bar d)$ can be expressed as
\begin{equation}
d\hat\sigma_D=\sum_{{\bf c},{\bf s},q}d\hat\sigma_{(c\bar q)^{\bf
c}_{\bf s}}~\rho((c\bar q)^{\bf c}_{\bf s}\rightarrow D).
\end{equation}
Where ${\bf c}$ and ${\bf s}$ denote the color and angular momentum
quantum numbers of the $(c\bar q)$ state, and $\rho((c\bar q)^{\bf
c}_{\bf s}\rightarrow D)$ is the non-perturbative parameter
characterizing the probability for the $(c\bar q)_{\bf s}^{\bf c}$
to evolve into a state including the $D$ meson.

For ${(c\bar q)^{\bf c}_{\bf s}}$ with various antiquark $\bar q$,
spin (${\bf s}$) and color (${\bf c}$), there may be a number of
$\rho$ parameters. However, the number can be greatly reduced in
terms of symmetries and some approximations. First of all, according
to the color-factor of the processes, the cross sections for the
production via color-octet and color-singlet $(c\bar q)$ differ by a
single factor ${1\over 8}$, therefore we can express the cross
section for $D$-meson production as follows:
\begin{equation}
d\hat\sigma_D=\sum_{{\bf s},{\bf q}}d\hat\sigma_{(c\bar q)^{\bf
1}_{\bf s}}~\rho_{\mathrm{eff}}((c\bar q)_{\bf s}\rightarrow D),
\end{equation}
with the definition:
$$\displaystyle \rho_{\mathrm{eff}}((c\bar
q)_{\bf s}\rightarrow D)\equiv\rho((c\bar q)_{\bf s}^{\bf
1}\rightarrow D)+{1\over 8}\rho((c\bar q)_{\bf s}^{\bf 8}\rightarrow
D)\,.$$ If the produced $D$-meson has different flavor from that of
the quark pair $(c\bar q)$, that means the quark pair $(c\bar q)$
must emit a flavored object (such as a pion etc) in the meantime
forming the $D$ meson, e.g., $(c\bar u)\rightarrow D^++\pi^-$, then
the relevant $ \rho_{\mathrm{eff}}$ will be relatively suppressed in
the large $N_c$ limit.
As in Refs.~\cite{br02c,br02prl}, we neglect such transitions from
the quark pair $(c\bar q)$ to a different flavored $D$-meson.
Furthermore, SU(3) light quark flavor symmetry indicates
$\rho_{\mathrm{eff}}((c\bar u)_s\rightarrow D^0)\simeq
\rho_{\mathrm{eff}}((c\bar d)_s\rightarrow D^+)$. As discussed in
Ref.~\cite{br02c}, heavy quark spin symmetry implies
\begin{equation}
\rho_{\mathrm{eff}}((c\bar q)_{^1S_0}\rightarrow D(c\bar
q))=\rho_{\mathrm{eff}}((c\bar q)_{^3S_1}\rightarrow D^*(c\bar
q))\,.
\end{equation}
Thus, only two independent parameters are left:
\begin{eqnarray}
\!\!\!\rho_{\mathrm{sm}}&\equiv&\rho_{\mathrm{eff}}((c\bar
d)_{^1S_0}\rightarrow
D^+)=\rho_{\mathrm{eff}}((c\bar d)_{^3S_1}\rightarrow D^{*+})\nonumber\\
&=&\rho_{\mathrm{eff}}((c\bar u)_{^1S_0}\rightarrow
D^0)=\rho_{\mathrm{eff}}((c\bar u)_{^3S_1}\rightarrow D^{*0}), ~~~~
\end{eqnarray}
\begin{eqnarray}
\!\!\!\rho_{\mathrm{sf}}&\equiv&\rho_{\mathrm{eff}}((c\bar
d)_{^1S_0}\rightarrow
D^{*+})=\rho_{\mathrm{eff}}((c\bar d)_{^3S_1}\rightarrow D^{+})\nonumber\\
&=&\rho_{\mathrm{eff}}((c\bar u)_{^1S_0}\rightarrow
D^{*0})=\rho_{\mathrm{eff}}((c\bar u)_{^3S_1}\rightarrow D^{0}).
~~~~
\end{eqnarray}

Now $\delta S^-_{\mathrm{HR}}(\xi,Q^2)$ caused by HQR, according to
Eq.(\ref{eq:SAHR}) can then be evaluated by means of
Eqs.(\ref{eq:M21},\ref{eq:M22}) with the auxiliary parameters
$\rho_{\mathrm{sm}}$ and $\rho_{\mathrm{sf}}$ which are determined
from relevant experiments. For the target to be proton, the HQR
correction of Eq.(\ref{eq:SAHR}) can be expressed as
\begin{eqnarray}\label{eq:P-SHQR}
\delta S^-_\mathrm{HR}(\xi,Q^2)=&&
 {\pi r_w^2\xi\over G_F^2  f_c B_{ef}|V_{cs}|^2}
\nonumber\\
\times\int dx&&\Big\{\big[d_v(x,\mu^2)Br_{D^+}+u_v(x,\mu^2)Br_{D^0}\big]\nonumber\\
\times&&\Big({d^2\hat{\sigma}_{(c\bar q)_{^1S_0}}\over d\xi
dQ^2}\rho_{sm}+{d^2\hat{\sigma}_{(c\bar
q)_{^3S_1}}\over d\xi dQ^2}\rho_{sf}\Big ) \nonumber\\
+&&
\big[d_v(x,\mu^2)Br_{D^{*+}}+u_v(x,\mu^2)Br_{D^{*0}}\big]\nonumber\\
\times&&\Big({d^2\hat{\sigma}_{(c\bar q)_{^1S_0}}\over d\xi
dQ^2}\rho_{sf}+{d^2\hat{\sigma}_{(c\bar q)_{^3S_1}}\over d\xi
dQ^2}\rho_{sm}\Big )\Big\}\,,
\end{eqnarray}
where the subprocess cross sections $d\hat{\sigma}_{(c\bar
q)_{^1S_0}}$ and $d\hat{\sigma}_{(c\bar q)_{^3S_1}}$ are related to
the `averaged and summed' squared amplitudes of
Eq.~(\ref{eq:M21},\ref{eq:M22}) and are independent of quark flavor
$q$. And for the nucleus target with proton number $P$ and neutron
number $N$, the HQR correction can be expressed as
\begin{eqnarray}\label{eq:A-SHQR}
\delta S^-_\mathrm{HR}(\xi,Q^2)=&&
 {\pi r_w^2\xi\over G_F^2  f_c B_{ef}|V_{cs}|^2}
\nonumber\\
\times\int dx&&\Big\{\Big[{P\,d_v(x,\mu^2)+N\,u_v(x,\mu^2)\over
P+N}Br_{D^+}\nonumber\\
+&&{P\,u_v(x,\mu^2)+N\,d_v(x,\mu^2)\over
P+N}Br_{D^0}\Big] \nonumber\\
\times&&\big({d^2\hat{\sigma}_{(c\bar q)_{^1S_0}}\over d\xi
dQ^2}\rho_{sm}+{d^2\hat{\sigma}_{(c\bar
q)_{^3S_1}}\over d\xi dQ^2}\rho_{sf}\big ) \nonumber\\
+&& \Big[{P\,d_v(x,\mu^2)+N\,u_v(x,\mu^2)\over
P+N}Br_{D^{*+}}\nonumber\\
+&&{P\,u_v(x,\mu^2)+N\,d_v(x,\mu^2)\over
P+N}Br_{D^{*0}}\Big]\nonumber\\
\times&& \big({d^2\hat{\sigma}_{(c\bar q)_{^1S_0}}\over d\xi
dQ^2}\rho_{sf}+{d^2\hat{\sigma}_{(c\bar q)_{^3S_1}}\over d\xi
dQ^2}\rho_{sm}\big )\Big\}\,.
\end{eqnarray}

For a quantitative estimate, in the following, we calculate the
$\delta S^-_{\mathrm{HR}}(\xi,Q^2)$ for isoscalar target ($P=N$),
with the parameters $\rho_{\mathrm{sm}}=0.15$,
$\rho_{\mathrm{sf}}=0$, which are taken from the extraction from
experimental charm photoproduction asymmetry by Ref.\cite{br02c}. In
this case, the HQR correction of Eq.(\ref{eq:A-SHQR}) can be
simplified,
\begin{eqnarray}\label{eq:iso-sf0-SHQR}
\delta S^-_\mathrm{HR}(\xi,Q^2)&=&
 {\pi r_w^2\xi\rho_{sm}\over G_F^2  f_c B_{ef}|V_{cs}|^2}
\int dx{d_v(x,\mu^2)+\,u_v(x,\mu^2)\over 2}\nonumber\\
&&\times \Big[(Br_{D^+}+Br_{D^0}) {d^2\hat{\sigma}_{(c\bar
q)_{^1S_0}}\over d\xi dQ^2}\nonumber\\
&&+ (Br_{D^{*+}}+Br_{D^{*0}}) {d^2\hat{\sigma}_{(c\bar
q)_{^3S_1}}\over d\xi dQ^2}\Big]\,.
\end{eqnarray}

In fact, for calculating Eq.(\ref{eq:iso-sf0-SHQR}) we need to carry
out a three-dimensional integration numerically. In the C.M. frame
of the subprocess, $\xi$ and $Q^2$ are related to the energy $k_0$
and the angle $\theta_1$ of the outgoing muon (from $\nu_\mu$ or
$\bar\nu_\mu$ vertex) relative to the incident direction:
\begin{equation}
\xi={x[{\sqrt s}k_0(1-\cos\theta_1)+m_c^2]\over s-{\sqrt
s}k_0(1+\cos\theta_1)}
\end{equation}
\begin{equation}
Q^2={\sqrt s}k_0(1-\cos\theta_1),
\end{equation}

Bearing the NuTeV dimuon experiment in mind, the incident energy of
neutrino or anti-neutrino is taken to be $E_\nu=160$ GeV, which is
approximately averaged value of the experiment. Furthermore, for our
numerical calculation, charm mass $m_c$ is fixed with the value
$1.5$ GeV, the coupling constant $\alpha_s(\mu)$ is running as
specified in CTEQ6L\cite{CTEQ6} and the parton distribution
functions of the nucleon are taken from CTEQ6L\cite{CTEQ6} too. The
branching ratio for the muonic decay of $D$ meson and the CKM matrix
elements are taken to be the central values in Ref.\cite{PDG06}, and
the $B_{ef}$ is taken to be the central value of
$B_{c\rightarrow\mu^+X}$ in Ref.\cite{PDG06}. Since the two opposite
charged muons in NuTeV experiment are required to have energy
greater than 5 GeV, so we make a cut for the produced $\mu$ and the
$D$ meson accordingly.

\begin{figure}
{\includegraphics[width=1.05\columnwidth]{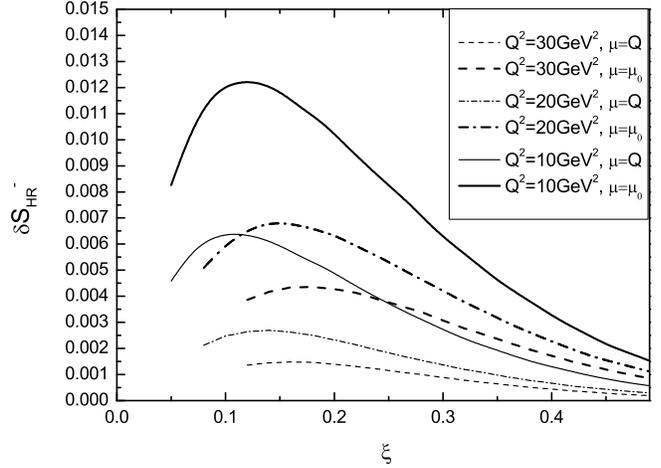}}
\caption{$\delta S^-_{\mathrm{HR}}$ as a function of $\xi$ for
$Q^2=10$~GeV$^2$ (solid lines), $Q^2=20$~GeV$^2$ (dash-dotted
lines), and $Q^2=30$~GeV$^2$ (dashed lines). Thick lines are results
for $\mu=\mu_0\equiv\sqrt{p_{c\perp}^2+m_c^2}$ and thin lines are
for $\mu=Q$.}
\end{figure}

The obtained result of $\delta S^-_{\mathrm{HR}}$ as a function of
$\xi$ for $Q^2=10$~GeV$^2$ (solid lines), $Q^2=20$~GeV$^2$
(dash-dotted lines), and $Q^2=30$~GeV$^2$ (dashed lines) is shown in
FIG.4. Since the calculation is of LO for the direct $D$-meson
production, there are theoretical uncertainties, such as that from
the energy scale $\mu$ of perturbative QCD, so to see the
uncertainty, we calculate $\delta S^-_{\mathrm{HR}}$ with two types
of choices about $\mu$. The thick lines in the figure present the
results where the factorization scale $\mu$ is taken to be
$\mu=\sqrt{p_{c\perp}^2+m_c^2}\equiv\mu_0$, where $p_{c\perp}$ is
the transverse momentum of the produced $D$ meson to the direction
of the $W$ boson in the nucleon rest frame. It is an analogous
choice to that in Ref.~\cite{br02c}, where its relevant charm
photoproduction is calculated under the factorization scale
$\mu=\sqrt{p_{\perp}^2+m_c^2}$ with $p_{\perp}$ being the transverse
momentum of the produced $D$ to the incident photon direction. The
results in an alternative scale choice $\mu=\sqrt{Q^2}=Q$, are also
shown by the thin lines in FIG.4. One fact that should be noted here
is that the cross section from the HQR process decreases very slowly
with the increase of the energy cut of the produced $D$ meson
$E_{\mathrm{cut}}$, namely, the `recombination' is not suppressed by
the cut taken in experiments very much. That can be understood by
the fact that the difference of the direct production of $D$-meson
by the HQR is related to the subprocess with valance quark inside a
nucleon (Eq.(\ref{eq:SAHR})) so that the $D$-meson relevant to the
difference can carry comparatively high energy (momentum) that
escapes from the cut quite a lot.

From FIG.4 one can see that at fixed $Q^2$ in each case, $\delta
S^-_{\mathrm{HR}}(\xi)$ peaks in the region $\xi=0.1-0.2$, over the
peak $\delta S^-_{\mathrm{HR}}$ decreases with $Q^2$ increases, and
the results with factorization scale $\mu=Q$ are smaller than those
with $\mu=\mu_0$. The uncertainty from the choice of the
factorization scale $\mu$ can also be seen when the scale
$\mu=\mu_0$ is varied by a factor of 2: the results become nearly
trebles when $\mu=\mu_0/2$, and the results reduce nearly by half
when $\mu=2\mu_0$. Generally the uncertainty may be suppressed by
NLO calculation, but we leave the study beyond the present
calculations.

\begin{figure}
{\includegraphics[width=1.05\columnwidth]{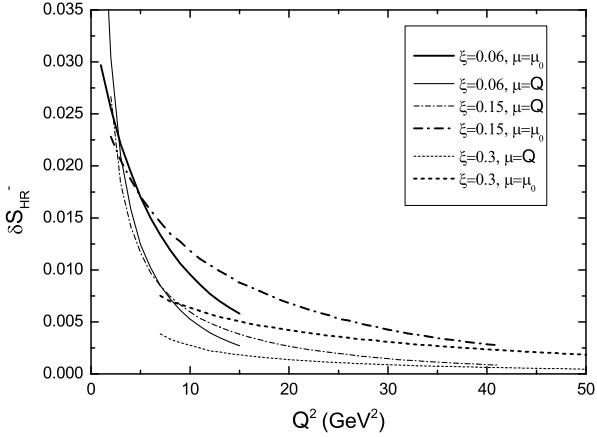}}\vspace{5mm}
 \caption{$\delta S^-_{\mathrm{HR}}$ as a function of $Q^2$ for
 $\xi=0.06$ (solid lines), $\xi=0.15$ (dash-dotted lines),
 and $\xi=0.3$ (dashed
lines). Thick lines are results for
$\mu=\mu_0=\sqrt{p_{c\perp}^2+m_c^2}$ and thin lines are for
$\mu=Q$.}
\end{figure}

The behavior of $\delta S^-_{\mathrm{HR}}$ as functions of $Q^2$ is
shown in FIG.5: the solid lines are those for $\xi=0.06$, the
dash-dotted lines are those for $\xi=0.15$ and dashed lines are
those for $\xi=0.3$. The thick lines are results for factorization
scale $\mu=\mu_0$ and the thin lines are those for $\mu=Q$.

In our calculation, a colinear singularity may arise from the
strange quark propagator in diagrams FIG.2(b) and FIG.2(d) when the
strange quark mass $m_s$ is set to be zero. In the limit
$x_q\rightarrow 0$, the denominator of the propagator is $2k_s\cdot
l$, which can reach zero when $\bf k_s$ and $\bf l$ are in the same
direction. To avoid this singularity, we have taken the strange
quark mass $m_s$ equal to its current mass 90 MeV in our above
numerical calculations.

\begin{figure}\label{fig:ms}
{\includegraphics[width=1.05\columnwidth]{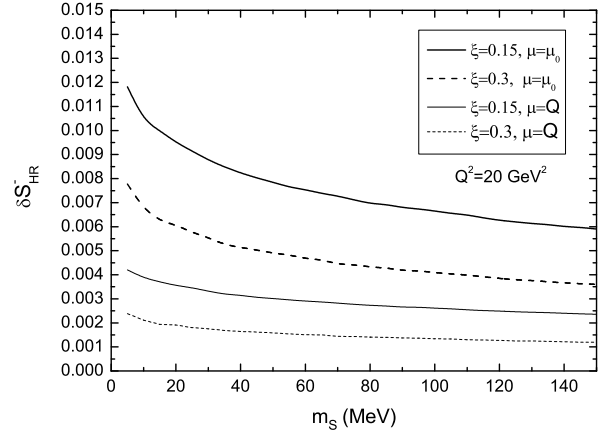}}
 \caption{$\delta S^-_{\mathrm{HR}}$ as a function of $m_s$ for
 $\xi=0.15$ (solid lines) and $\xi=0.3$ (dashed lines) at $Q^2=20 $GeV$^2$.
 A plot to show the resultant uncertainties from choosing $m_s$ in treatment of the collinear singularity.
 Thick lines are results for $\mu=\mu_0=\sqrt{p_{c\perp}^2+m_c^2}$
and thin lines are for $\mu=Q$.}
\end{figure}
There are some uncertainties in the treatment of this singularity.
To see this, in FIG.6, we show the results of $\delta
S^-_{\mathrm{HR}}$ as a function of $m_s$ for $Q^2=20$ GeV$^2$ at
two $\xi$ values for two choices of $\mu$. Generally, the results
decrease with the increase of $m_s$, and in the favored range 70-120
MeV for $m_s$ , the results may vary within 15\%. Lower $m_s$ range
shows greater $m_s$ dependence. There is another way to avoid the
singularity, i.e., to keep the $x_q$ suppressed terms in the s quark
propagator and take $x_q$ to be reasonable finite instead of zero.
When taking $x_q=1/6$ (approx light constitute mass over D meson
mass), the numerical results reduce nearly by half.

As indicated by the results above, when measuring the strange
distribution asymmetry inside a nucleon, one should consider the
correction $\delta S^-_{\mathrm{HR}}(\xi,Q^2)$ caused by HQR, which
is comparable to the existent measured value. For example, at
$Q^2=20$~GeV$^2$ (about averaged value in NuTeV experiment) and for
$\mu=\mu_0$, the HQR correction to strangeness asymmetry by
integrating $\delta S^-_{\mathrm{HR}}(\xi)$ over $\xi$ can be 0.002
approximately. To be comparison, the recent NLO analysis of the
NuTeV dimuon data~\cite{mason07} and the global
analysis~\cite{lai07} present the central value of the strangeness
asymmetry $S^-\approx0.002$. Thus, the HQR could not be negligible
in the extraction of the strangeness asymmetry.

The HQR correction may enhance the strangeness asymmetry by a larger
positive value, and large positive strangeness asymmetry can help to
explain the NuTeV anomaly~\cite{oln03,kre04,dm04,alw04,dxm04,wak04}.

The value of the parameter $\rho_{\mathrm{sm}}$ still has some
uncertainty as discussed in Refs.\cite{br02c,br02prl}, whereas, the
magnitude order of $\rho_{\mathrm{sm}}$ can not be changed and the
influence from HQR to the measurement of the nucleon strangeness
asymmetry could not be negligible. Moreover, more accurate $\rho$
parameters are needed not only for a better understanding of the HQR
effect but also for better determination of the strangeness
asymmetry inside a nucleon.

\section{Calculation of the LQF-P process}

The diagrams for the LQF-P process in $\nu_\mu$-induced CC DIS are
shown in FIG.3. The subprocess can contribute to the $\nu_\mu$- and
$\bar\nu_\mu$-induced dimuon cross section difference as shown in
Eq.(\ref{eq:csd-LQF-P}), and thus may also influence the measurement
of the strangeness asymmetry.

The calculation of the cross section $d\hat\sigma_{\bar cD}$ for
LQF-P process can be factorized into the convolution of the LO
subprocess cross section for light quark $\bar q$ production and the
fragmentation function $D_{\bar q}^D(z)$ of the light quark $\bar q$
into a $D$ meson:
\begin{equation}\label{hatcs-LQF}
d\hat\sigma_{\bar cD}=\sum_q\int_0^1 dz~d\hat\sigma_{\nu\bar
u\rightarrow\mu^-\bar q}D_{\bar q}^D(z)\,,
\end{equation}
where $\bar q$ could be $\bar d$ or $\bar s$, with $D$ being $D^{+}$
or $D_s^{+}$ or $D^{*+}$ or $D_s^{*+}$. With $CP$ transformation and
SU(3) flavor symmetry, only two independent light-quark
fragmentation functions remain:
\begin{eqnarray}
&D_q(z)\equiv D_{\bar d}^{D^+}(z)=D_{\bar s}^{D_s^+}(z)\,,\nonumber\\
&D_q^*(z)\equiv D_{\bar d}^{D^{*+}}(z)=D_{\bar s}^{D_s^{*+}}(z)\,.
\end{eqnarray}


The fragmentation functions are calculable with
pQCD\cite{chang92,br93prl,br93c,br93bc,chen93}. For $\bar
q\rightarrow D(c\bar q)$, the fragmentation function can be
expressed as~\cite{br93bc}
\begin{eqnarray}\label{eq:MM22}
&&D_{\bar q}^D(z) ={1\over 16\pi^2}\nonumber\\
&&\times\int ds ~\theta(s-{(m_q+m_c)^2\over z}-{m_c^2\over
1-z})\lim_{q^0\rightarrow\infty}{|M|^2\over|M_0|^2},
\end{eqnarray}
where $M$ and $M_0$ are amplitudes for the $D$ production and the LO
on shell $\bar q$ production respectively; Let $P_D=p_q+p_c$ and
$k_c$ denote the momenta of the produced $D$ meson and the $\bar c$
quark respectively. Then $q=P_D+k_c$ is their total momentum, and
$s=q^2$.  The variable $z$ is defined by $z={(P_D^0+P_D^3)/
(q^0+q^3)}$ in a frame, where $q=(q^0,0,0,q^3)$. In axial gauge, the
amplitude corresponding to the diagram FIG.3(b) is suppressed and
can be neglected~\cite{chen93}. Thus only diagram FIG.3(a)
contributes. The formation of the bound state can be described by
the B-S wave functions, for the production of $^1S_0$ state
color-singlet $D$ meson,
\begin{equation}
\chi_p(^1S_0)={R(0)\over 3\sqrt {3\pi M_D}}
\gamma_5(P_D\!\!\!\!\!\!\!/~~+M_D),
\end{equation}
and for the production of $^3S_1$ state color-singlet $D$ meson,
\begin{equation}
\chi_p(^3S_1)={R(0)\over 3\sqrt {3\pi M_D}}
~\epsilon\!\!\!/(P_D\!\!\!\!\!\!\!/~~+M_D)\,,
\end{equation}
which should appear in the amplitude $M$ of Eq.(\ref{eq:MM22}). Here
$R(0)$ is the non-relativistic radial wave function at the origin
for the $D$ meson, and $\epsilon$ is the polarization vector of the
$^3S_1$ state $D$ meson.

The fragmentation function for light quark into $^1S_0$ state $D$
meson is given by~\cite{br93bc}
\begin{eqnarray}\label{FFS01}
D_q(z)&=&{2\alpha_s(2m_c)^2|R(0)|^2\over 81\pi m_c^3}{r
z(1-z)^2\over[1-(1-r)z]^6}\nonumber\\
&&\times\big[6-18(1-2r)z+(21-74r+68r^2)z^2\nonumber\\
&&-2(1-r)(6-19r+18r^2)z^3\nonumber\\
&&+3(1-r)^2(1-2r+2r^2)z^4\big],
\end{eqnarray}
where $r=m_c/M_D$. And the fragmentation function for light quark
into $^3S_1$ state  $D$ meson is given by~\cite{br93bc}
\begin{eqnarray}\label{FFS13}
D_q^*(z)&=&{2\alpha_s(2m_c)^2|R(0)|^2\over 27\pi m_c^3}{r
z(1-z)^2\over[1-(1-r)z]^6}\nonumber\\
&&\times\big[2-2(3-2r)z+3(3-2r+4r^2)z^2\nonumber\\
&&-2(1-r)(4-r+2r^2)z^3\nonumber\\
&&+(1-r)^2(3-2r+2r^2)z^4\big].
\end{eqnarray}

The value of $R(0)$ can be estimated from the pseudoscalar meson
decay constant $f_D$ through the relation
\begin{equation}
R(0)=\sqrt{\pi M_D\over 3} f_D.
\end{equation}
We take the central values from Ref.\cite{PDG06} for $M_{D^+}$ and
$f_{D^+}$ in calculating $R(0)$, and obtain $R(0)=0.31$ GeV$^{3\over
2}$. We take one-loop $\alpha_s$ with $\Lambda=326$ MeV for 4
flavors as in CTEQ6L, and obtain $\alpha_s(2m_c)=0.255$ for
$m_c=1.5$ GeV. The $r$ value in Eqs.(\ref{FFS01},\ref{FFS13}) is
evaluated by taking $M_D=1.87$ GeV and $M_{D^*}=2.01$
GeV\cite{PDG06}. The fragmentation functions $D_q(z)$ and $D_q^*(z)$
calculated with Eqs.(\ref{FFS01},\ref{FFS13}) are shown in FIG.7. If
integrating $D_q(z)$ and $D_q^*(z)$ over $z$, then $D_q\equiv\int
D_q(z)dz=2.01\times 10^{-5}$ and $D_q^*\equiv\int
D_q^*(z)dz=1.77\times 10^{-5}$ are obtained.
\begin{figure}
{\includegraphics[width=\columnwidth]{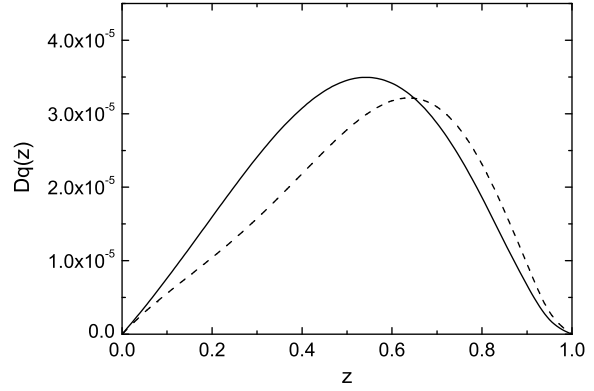}}
\caption{Fragmentation functions $D_q(z)$ and $D_q^*(z)$ for the
LQF-P process. The solid line denotes $D_q(z)$, and the dashed line
denotes $D_q^*(z)$.}
\end{figure}


With the formulas for factorization and the fragmentation functions
Eqs.(\ref{eq:csd-LQF-P},\ref{hatcs-LQF},\ref{FFS01},\ref{FFS13}),
for isoscalar target, the `correction' $\delta S^-_{\mathrm{LQF-P}}$
from LQF-P to measured strangeness asymmetry can be computed:
\begin{eqnarray}
\delta S^-_{\mathrm{LQF-P}}&\approx& \int dx dz {\pi r_w^2x\over
G^2_F |V_{cs}|^2f_cB_{ef}}{u_v(x)+d_v(x)\over 2}\nonumber\\
&\times&\Big\{{d\hat\sigma_{\nu\bar u\rightarrow\mu^-\bar d(\bar
s)}\over dQ^2}[D_q(z)Br_{D^+}+D_q(z)^*Br_{D^{*+}}]\Big\} \nonumber\\
&\approx& { 1\over 2|V_{cs}|^2B_{ef}}\{
Y(D_qBr_{D^+}+D_q^*Br_{D^{*+}})\}\,.
\end{eqnarray}
Here $Y=\int x[u_v(x)+d_v(x)](1-y)^2 dx$ with $y=Q^2/xS<1$. $Y=0.16$
is evaluated from the CTEQ6L parton distributions at $Q^2=20$
GeV$^2$ and $E_\nu=160$ GeV, and the muonic decay rates are taken to
be the central value from Ref.~\cite{PDG06}. With the parameters
given above, the final result of $\delta S^-_{\mathrm{LQF-P}}$ is
obtained:
\begin{eqnarray}\label{FFS113}
\delta S^-_{\mathrm{LQF-P}}\approx 0.53\times 10^{-5}\,.
\end{eqnarray}
Such a `correction' to the strangeness asymmetry from the LQF-P
process is much smaller than that measured. Thus, LQF-P gives little
influence in the extraction of the nucleon strangeness asymmetry.

We should note here that as pointed out in Section II, LQF-P in
neutrino and anti-neutrino DIS may generate not only the oppositely
charged dimuon events, but also the same charged dimuon events and
trimuon events (not from direct $D$-meson production), thus the fact
that in neutrino and anti-neutrino DIS experiments either the same
charged dimuon events or trimuon events are very rare is consistant
with the small value of $\delta S^-_{\mathrm{LQF-P}}$ as shown in
Eq.(\ref{FFS113}).

\section{conclusions}

The measurement of the nucleon strangeness asymmetry is important
for the study of nucleon structure and certain related phenomenon.
The cross section difference between the dimuon production from
neutrino and anti-neutrino DIS is sensitive observable to the
strange distribution asymmetry. Whereas in this work, we point out
two types of direct charmed meson production, i.e., HQR and LQF-P at
order $\alpha_s^2$, which also contribute to oppositely charged
dimuon production. These processes are not included in the
experimental analysis, therefore we further study their influence to
the measurements of the nucleon strangeness asymmetry. With
quantitative calculations in terms of pQCD, we find that HQR affects
the extraction of the strange distribution asymmetry with a positive
`correction' $\delta S^-_{\mathrm{HR}}(\xi,Q^2)$, and the
`correction' can be so large as $\delta S^-_{\mathrm{HR}}\sim
10^{-3}$. For the other one, LQF-P provides a small `correction' to
the measurement, that is of the order $\lesssim 10^{-5}$, so that it
can be neglected. The influence of HQR to the measurement of the
nucleon strangeness asymmetry from neutrino and anti-neutrino DIS
can not be negligible, that may provide a positive correction to the
present value of the strangeness asymmetry, and is also helpful to
explain the NuTeV anomaly. We think that a reanalysis of the strange
distribution asymmetry with consideration of the direct-D production
from HQR process is needed to improve the value of the nucleon
strangeness asymmetry.

$D$-meson directly produced in neutrino and anti-neutrino deep
inelastic scattering (DIS) contributes to oppositely charged
dimuon events, 
which has not been considered in experimental analysis so far. Hence
we conclude that in determining the strangeness asymmetry in a
nucleon via measuring the dimuon production in neutrino and
anti-neutrino DIS, the contribution from HQR production of the
direct $D$-meson should be deducted precisely.

{\bf Acknowledgments: } We are very grateful to C.-H. Chang for
suggestions and contributions.  We also thank K.-T. Chao, Y.-Q.
Chen, Y.-J. Gao and Y. Jia for helpful discussions. This work is
partially supported by National Natural Science Foundation of China
(Nos.~10721063, 1057-5003, 10528510), by the Key Grant Project of
Chinese Ministry of Education (No.~305001), by the Research Fund for
the Doctoral Program of Higher Education (China).

\end{document}